\documentclass[journal]{IEEEtran}

\usepackage{amsmath}
\usepackage{amssymb}
\usepackage{latexsym}
\usepackage{amsmath}
\usepackage{amsthm}
\usepackage{amssymb}
\usepackage{enumerate}
\usepackage{latexsym}
\usepackage{mathtools}
\usepackage{amsfonts}
\usepackage{setspace}
\usepackage{cite}
\ifCLASSINFOpdf
\else
\fi
\begin{document}
%
\title{\bf Coordinate-ordering-free Upper Bounds for Linear Insertion-Deletion Codes}
%
%
%

\author{Hao Chen 
\thanks{Hao Chen  is with the College of Information Science and Technology/Cyber Security, Jinan University, Guangzhou, Guangdong Province, 510632, China,  e-mail: (haochen@jnu.edu.cn).}
\thanks{Manuscript received July 6, 2021; revised February. 2, 2022, accepted April 11, 2022. This research was supported by NSFC Grant 62032009.}}

%
%

\markboth{Journal of \LaTeX\ Class Files,~Vol.~14, No.~8, August~2015}%
{Shell \MakeLowercase{\textit{et al.}}: Bare Demo of IEEEtran.cls for IEEE Journals}
%



\maketitle

\begin{abstract}
In this paper we prove several coordinate-ordering-free upper bounds on the insdel distances of linear codes. Our bounds are stronger than some previous known bounds. We apply these upper bounds to AGFC codes from some cyclic codes and one algebraic-geometric code with any rearrangement of coordinate positions. A strong upper bound on the insdel distances of Reed-Muller codes with the special coordinate ordering is also given.
\end{abstract}


%
\IEEEpeerreviewmaketitle

\section{Introduction}

It has been a long-standing challenge to deal efficiently with synchronization errors, i.e., insertions and deletions, see \cite{L65,L1992,L2002}. The early motivation to study the common subsequence also came from its biological application, see \cite{BM,CS,SK}. The insertion-deletion codes were proposed to deal with synchronization errors and have wide applications in racetrack memory error-corrections, language processing, data analysis and DNA storage, see \cite{SK,Chee,Brill,Xu,JHSB17,LSWY}. There have been continuous efforts to construct codes correcting one or two deletion/insertion errors, see \cite{BGZ,GS,SRB,SB,NJAS,SWGY,VT}.  We refer to \cite{L65,VT,T84,L2002,DM,DA,AGFC,M2009,MBT, SZ99, BGMO16,GHS} for the historic development of insertion-deletion error-correcting codes. For the recent breakthroughs and constructions we refer to \cite{HS17,HS18,HSS18,HS20,CJLW18,CHLSW19,GHS,CGHL21,GS,SSBD,SWGY,SRB,SZ99,SB,SWWY,LSWY,TFV} and a nice latest survey \cite{HS21}.  Efficient coding attaining the near-Singleton optimal rate-distance tradeoff was achieved in \cite{HS17,HS18}.\\

For a vector ${\bf a}=(a_1, \ldots, a_n) \in {\bf F}_q^n$, the support of ${\bf a}$ is $$supp({\bf a}) =\{i_h:a_{i_h} \neq 0\}.$$ The Hamming weight $wt({\bf a})$ of ${\bf a}$ is the number of coordinate positions in its support. The Hamming distance $d_H({\bf a}, {\bf b})$ between two vectors ${\bf a}$ and ${\bf b}$ is defined to be the Hamming weight of ${\bf a}-{\bf b}$. For a linear code ${\bf C} \subset {\bf F}_q^n$ of dimension $k$, its minimum Hamming distance $d_H$ is the minimum of Hamming distances $d_H({\bf a}, {\bf b})$ between any two different codewords ${\bf a}$ and ${\bf b}$ in ${\bf C}$. It is well-known that the minimum Hamming distance of a linear code ${\bf C}$ is the minimum Hamming weight of its non-zero codewords. The famous Singleton bound $d_H \leq n-k+1$ on the minimum Hamming distance $d_H$ of an $[n,k,d_H]_q$ code is the basic upper bound for linear error-correcting codes in the Hamming-metric. A linear code attaining this bound is called a MDS (maximal distance separable) code. A long-standing conjecture in the theory of linear Hamming error-correcting codes is the main conjecture of the MDS codes, which asserts that the length of a linear MDS code over ${\bf F}_q$ can not be bigger than $q+1$ except some obvious trivial cases, we refer to \cite{Roth}.\\

The support of a linear sub-code $D \subset {\bf C}$ is $$supp(D)=\{1 \leq i \leq n: x_i \neq 0: \exists x=(x_1,\ldots,x_n) \in D\},$$ that is, the support of a linear sub-code $D$ is the non-zero coordinate positions of all codewords in $D$. The $r$-th generalized Hamming weight $d_r$ for $1\leq r \leq k$ is defined to be the minimum of the number of support positions of arbitrary $r$ dimension sub-codes. Hence $d_1$ is the minimum Hamming distance. It is clear that $d_1<d_2 < \cdots <d_k$ and the generalized Singleton bound $d_r \leq n-k+r$ is satisfied for a linear $[n,k]_q$ code. On the other hand the Plotkin bound on the generalized Hamming weights $d_r \leq [\frac{n(q^r-1)q^{k-r}}{q^k-1}]$ was proved in \cite{TV1}, see Theorem 3.1 \cite{TV1}. The generalized Hamming weights have been calculated for many linear codes, for example, see \cite{Wei,HP98}.\\

We define the partial ranks of a linear code as the dimensions of the projection codes to subsets of coordinate positions. For a linear $[n,k]_q$ code ${\bf C} \subset {\bf F}_q^n$ and the subset of coordinate positions $S=\{i_1, \ldots, i_h\} \subset \{1,\ldots,n\}$, the natural mapping $\Phi_S: {\bf C} \longrightarrow {\bf F}_q^h$ is defined by $\Phi_S({\bf x})=(x_{i_1},\ldots,x_{i_h})$, where ${\bf x}=(x_1,\ldots,x_n) \in {\bf C}$.  We define the partial rank function of the code ${\bf C}$ at ${\bf x}$ as $$rank({\bf x}, {\bf C})=\dim (\Phi_{supp({\bf x})}({\bf C})).$$\\

The insdel distance $d_{insdel}({\bf a}, {\bf b})$ between two vectors ${\bf a}$ and ${\bf b}$ in ${\bf F}_q^n$ is the number of insertions and deletions which are needed to transform ${\bf a}$ into ${\bf b}$. Actually it was proved in \cite{Duc21} \cite{HS17} that $$d_{insdel}({\bf a}, {\bf b})=2(n-l),$$ where $l$ is the length of a longest common subsequence of ${\bf a}$ and ${\bf b}$. This insdel distance $d_{insdel}$ is indeed a metric on ${\bf F}_q^n$. It is clear $d_{insdel}({\bf a}, {\bf b}) \leq 2d_H({\bf a}, {\bf b})$ since  $l \geq n-d_H({\bf a}, {\bf b})$ is valid for arbitrary two different vectors ${\bf a}$ and ${\bf b}$ in ${\bf F}_q^n$. The insdel distance of a code ${\bf C} \subset {\bf F}_q^n$ is the minimum of the insdel distances of all different two codewords in this code. Hence we have the direct upper bound $d_{insdel}({\bf C}) \leq 2d_H({\bf C})$, and the direct Singleton upper bound on the insdel distance of a linear $[n,k]_q$ code $$d_{insdel} \leq 2(n-k+1),$$   see \cite{BGMO16,HS17}. The relative insdel distance is defined as $\delta=\frac{d_{insdel}}{2n}$ since $d_{insdel}$ takes non-negative integers up to $2n$. From the Singleton bound $d_{insdel} \leq 2(n-k+1)$ it follows immediately $$R+\delta \leq 1.$$ For insertion-deletion codes the ordering of coordinate positions strongly affects the insdel distances. In this paper we give some upper bounds for insdel distances of linear codes which are valid for any fixed ordering of coordinate positions. \\

Most recent new constructions of efficient insertion-deletion codes are not linear, except the construction in \cite{CGHL21}. It is obvious that linear codes have advantages in both theory and practice because of their compact representations and highly efficient encoding. There are a lot of nice linear Hamming error-correcting codes from the algebraic coding technique. On the other hand for linear codes very few upper or lower bounds on their inedel distances have been known. In \cite{L65,VT,T84,DM,DA,AGFC,MS07} the insertion-deletion error-correcting capabilities of cyclic codes, Reed-Muller codes and Reed-Solomon codes were analysed. A better understanding of the insertion-deletion error-correcting capabilities of linear codes is needed.\\

For Hamming error-correcting codes,  a basic result about linear codes is the Gilbert-Varshamov bound can be achieved by a non-constructive counting proof. A Gilbert-Varshamov bound for general (not linear) insertion-deletion codes was proved in \cite{LTX} Proposition 7. In paper \cite{AGFC} it was proved that a linear code that can correct even a single deletion are limited to have information rate at most $\frac{1}{2}$. The explicit construction of binary linear code sequence with  the rate $0<R<\frac{1}{2}$ and correcting $\delta>0$ fraction of insdel errors was given in \cite{CGHL21}. In \cite{CGHL21} Section 5 the asymptotic half-Singleton bound was proved in Corollary 5.2. Their bound can be restated as $d_{insdel} \leq \max\{2(n-2k+2),2\}$, see Section 2 below. It was also proved in \cite{CGHL21}, Section 5 that there exists no sequence of linear $[n(t),  k(t)]_q$ codes over ${\bf F}_q$ with insdel distances $d(t)$, $t=1,2,\ldots,$ and the code length $n(t)$ goes to the infinity, such that $$R=\lim_{t \longrightarrow \infty} \frac{k(t)}{n(t)} \geq \frac{1}{2},$$ and $$\delta=\lim_{t \longrightarrow \infty} \frac{d(t)}{2n(t)}>0.$$ Their proof is based on their half-Singleton bound or the half-Plotkin bound in \cite{CGHL21}.\\

Let ${\bf F}_q$ be an arbitrary finite field, $P_1,\ldots,P_n$ be $n \leq q$ elements in ${\bf F}_q$. The Reed-Solomon codes $RS(n,k)$ is defined by $$RS(n,k)=\{(f(P_1),\ldots,f(P_n)): f \in {\bf F}_q[x],\deg(f) \leq k-1\}.$$ This is a $[n,k,n-k+1]_q$ linear MDS codes from the fact that a degree $\deg(f) \leq k-1$ polynomial has at most $k-1$ roots. It was proved in \cite{MS07} that for Reed-Solomon codes of length $n \geq 3$ and dimension $2$ over large prime finite fields ${\bf F}_p$ the insdel distance can never meet the above direct Singleton bound. This was improved recently in a result of Duc, Liu, Tjuawinata, Xing proved in \cite{Duc21}.  They proved that the insdel distances of $k$ dimension Reed-Solomon codes has to satisfy  $$d_{insdel}\leq 2n-2k$$ if $n>k>1$ and $q>n^2$. This Singleton type bound $$d_{insdel} \leq 2n-2k$$ was proved further for a general linear $[n,k]_q$ code over an arbitrary finite field ${\bf F}_q$ satisfying $n>k \geq2$ in \cite{CZ21}. For the dimension $k=2$ case optimal Reed-Solomon codes attaining this bound were constructed in \cite{Duc21,CZ21}. However the lengths of these two dimensional optimal codes are very small comparing with the size $q$ of the field. As the main conjecture of the linear MDS codes for the Hamming metric, the longest possible length of "optimal" linear insertion-deletion codes attaining the new Singleton type upper bound in \cite{CGHL21} and our this paper, if exist, is a very challenge problem. From the half-Singleton bound in \cite{CGHL21} we have the following upper bounds for the insdel distances of the Reed-Solomon codes. For an $[n,k,n-k+1]_q$ Reed-Solomon code satisfying $2k \geq n+1$ we have $d_{insdel} \leq 2$ from the half-Singleton bound. For a dimension $3$ Reed-Solomon code with the minimum Hamming distance $n-2$ its insdel distance satisfies $$d_{insdel} \leq 2n-8.$$\\

We give a new upper bound Theorem 2.1 on insdel distances of linear codes based on the positions of information free subsets.  The previous upper bounds $d_{insdel} \leq 2(n-k)$ in \cite{Duc21,CZ21} follows from our main result Theorem 2.1 immediately. In some cases the half-Singleton bound $d_{insdel} \leq 2(n-2k+2)$ follows from our main result Theorem 2.1. From our main result Theorem 2.1 we give a new upper bound on $d_{insdel}$ of a linear code which depends on the formation of minimum Hamming weight codewords in this linear code. Its strongest form is as follows. If there is a minimum Hamming weight codeword ${\bf x}$ with consecutive index support, then $$d_{insdel} \leq 2(d_H-rank({\bf x}, {\bf C})+1), $$ where $d_H$ is the minimum Hamming weight $wt({\bf x})$, $rank({\bf x}, {\bf C})$ is the dimension of the projection code to the support of ${\bf x}$. When the minimum Hamming distance of a linear code satisfying $d_H \leq n-2k+2$, our this bound is stronger than the half-Singleton bound and the direct bound $d_{insdel} \leq 2d_H$.\\

We apply our new bound to an algebraic-geometric code and some binary Reed-Muller codes. A strong coordinate ordering-depending upper bound on insdel distances of binary Reed-Muller codes is presented. From our upper bounds on insdel distances from partial ranks, we prove that with certain fixed coordinate ordering, the insdel distances of some binary Reed-Muller codes $RM(u,m)$ are at most $poly(m)$, which are quite smaller than their exponential Hamming distances $d_H=2^{m-u}\geq 2^{m/2}$, when $u$ is smaller and very close to $\frac{m}{2}$. This is much stronger than the direct bound $d_{insdel} \leq 2d_H$ and the half-Singleton bound $d_{insdel} \leq 2(n-2k+2)$. For Reed-Solomon codes if these upper bounds are attained, the lengths have to be very small. We speculate that Reed-Solomon codes and their generalizations algebraic-geometric codes are good candidates as linear codes with moderate good insertion-deletion error-correcting capabilities.\\

\section{Main results}

For a linear $[n,k]_q$ code ${\bf C} \subset {\bf F}_q^n$,  the subset $S\subset \{1,\ldots,n\}$ of $h$ coordinate positions is called an information free coordinate subset if  the natural projection $\Phi_S: {\bf C} \longrightarrow {\bf F}_q^h$ defined by $\Phi_S((c_1, \ldots, c_n))=(c_{i_1},\ldots, c_{i_h})$ is surjective. It is clear $h \leq k$. When $h=k$ this is the information set.\\

{\bf Theorem 2.1.} {\em Let ${\bf C} \subset {\bf F}_q^n$ be a linear $[n,k]_q$ code with an information free coordinate subset $S=\{i_1,\ldots,i_h\}$ of the cardinality $h \leq k$, where $1\leq i_1<i_2<\cdots<i_h\leq n$. If there exists a codeword ${\bf x} \in {\bf C}$ with $n-h-t$ zero coordinate positions in the range $[1,i_1-1]$ or $[i_h+1,n]$. Then the insdel distance of this code satisfies $$d_{insdel} \leq 2(t+1).$$}\\

\begin{proof}
We assume that $S=\{i_1,i_2,\ldots,i_h\}$ is an information free coordinate set of $h$ coordinate positions, where $i_1<i_2<\cdots<i_h$. Set $S'=\{1,2,\ldots,n\}-S$. The length $h$ and $n-h$ vectors located at the set $S$ and $S'$ of a vector ${\bf y} \in {\bf F}_q^n$ are denoted by ${\bf y}_S$ and ${\bf y}_{S'}$. The main point of the proof is as follows. Since the set $S$ is an information free coordinate subset, there exists a codeword with any given coordinate values in these $h$ coordinate positions of $S$. Then we can construct a codeword ${\bf a}$ and make that the common subsequence of ${\bf a}_S $ and ${\bf a}_S+{\bf x}_S$ has the length $h-1$. On the other hand since there are $n-h-t$ zero coordinate positions of ${\bf x}$ before or after this information free subset $S$. There is  a length $n-h-t$ common subsequence of ${\bf a}$ and ${\bf a}+{\bf x}$ in the coordinate positions $[1, i_1-1]$ and $[i_h+1, n]$. Then there is a long common subsequence in the codeword ${\bf a}$ and ${\bf a}+{\bf x}$ of the length at least $h-1+n-h-t=n-t-1$.\\

Let ${\bf x}=(x_1,x_2,\ldots,x_n)$ be the codeword described in the Theorem 2.1. From the condition that $S=\{i_1,\ldots, i_h\}$ is an information free coordinate set, since the mapping ${\bf \Phi}_S: {\bf C} \longrightarrow {\bf F}_q^h$ defined by $\Phi_S({\bf y})=(y_{i_1}, \ldots, y_{i_h})$, is surjective, we can find a codeword ${\bf a} \in {\bf C}$ satisfying that ${\bf a}_S=(a_{i_1},a_{i_2},\ldots,a_{i_{h-1}},a_{i_h})$ with the following coordinate values $$a_{i_2}=a_{i_1}-x_{i_2},$$ $$a_{i_3}=a_{i_1}-x_{i_2}-x_{i_3},$$ $$\cdots,$$ $$a_{i_h}=a_{i_1}-x_{i_2}-x_{i_3}-\cdots-x_{i_h}.$$ Here $a_{i_1}$ is an arbitrary element in ${\bf F}_q$. Then $${\bf a}_S+{\bf x}_S=(a_{i_1}+x_{i_1}, a_{i_1}, a_{i_1}-x_{i_2}, \ldots, a_{i_1}-x_{i_2}-\cdots-x_{i_{h-1}}),$$ and $${\bf a}_S=(a_{i_1}, a_{i_1}-x_{i_2},\ldots, a_{i_1}-x_{i_2}-\cdots-x_{i_{h-1}}, a_{i_1}-x_{i_2}-\cdots-x_{i_h}),$$ there is a length $h-1$ common subsequence in ${\bf a}_S$ and ${\bf a}_S+{\bf x}_S$.\\

This common subsequence of ${\bf a}_S$ and ${\bf a}_S+{\bf x}_S$ has their positions in the range $[i_1,i_h]$. Since there are $n-h-t$ zero coordinate positions of the codeword ${\bf x}$ in  $[1,i_1-1]$ and $[i_h+1,n]$, then there is a length $n-h-t$ common subsequence of ${\bf a}_{S'}+{\bf x}_{S'}$ and ${\bf x}_{S'}$  such that their positions are in $[1,i_i-1]$ and $[i_h+1,n]$.  Therefore we can patch the two common subsequences of lengths $n-h-t$ and $h-1$ without change the coordinate ordering.  The length of the common subsequence of ${\bf a}$ and ${\bf a}+{\bf x}$ is at least $h-1+n-h-t=n-t-1$. Then $d_{insdel}({\bf a}, {\bf x}+{\bf a})\leq 2(n-(n-t-1))=2(t+1)$. The conclusion follows directly.
\end{proof}

Actually Theorem 2.1 is general to include some previous upper bounds. First of all at arbitrarily given $H\geq k-1$ coordinate positions, there is a nonzero codeword vanishing at arbitrary $k-1$ coordinate positions among these $H$ positions, since for any $k-1$ columns in a generator matrix of this code, we can find an length $k$ vector orthogonal to these $k-1$ columns.  Then in the most general case when $i_1=1, i_n=n$, we can set $h=k, n-k-t=0$. The upper bound in Theorem 2.1 is $d_{insdel} \leq 2(n-k+1)$, which is the direct Singleton bound. When $k \geq 2$, it is clear that there are two linearly independent consecutive columns in the generator matrix, then $h=2, i_2=i_1+1$, and we can find a codeword which have $k-1$ zero positions outside the coordinate position set $\{i_1,i_1+1\}$. Then $n-2-t=k-1$, $t=n-k-1$, we have $d_{insdel} \leq 2(n-k)$ from Theorem 2.1.  Hence our main result Theorem 2.1 is much stronger than the previous upper bound $d_{insdel} \leq 2(n-k)$ in \cite{Duc21,CZ21}. In general if we can find consecutive linear independent $h \leq k$ columns in a generator matrix of this linear code, then $d_{insdel}\leq 2(n-h-k+2)$. In the case that there is an information set with consecutive coordinate positions, the half-Singleton bound $d_{insdel} \leq 2(n-2k+2)$ in \cite{CGHL21} follows from Theorem 2.1. Since arbitrary  $k$ columns in the generator matrix of an MDS $[n,k,n-k+1]_q$ code are linear independent, the half-Singleton bound of an MDS code follows from our main result Theorem 2.1.\\

The following result follows from Theorem 2.1 directly.\\

{\bf Corollary 2.1.} {\em For a linear $[n,k]_q$ code ${\bf C} \subset {\bf F}_q^n$ and any given non-zero codeword ${\bf x} \in {\bf C}$ with $S({\bf x})$ the smallest index and $L({\bf x})$ the largest index in its support, we have $$d_{insdel} \leq 2(L({\bf x})-S({\bf x})-\operatorname{rank}({\bf x},{\bf C})+2).$$ If $L=wt({\bf x})+S({\bf x})-1$, that is, $supp({\bf x})$ is a set of consecutive indices, then $$d_{insdel} \leq 2(wt({\bf x})-\operatorname{rank}({\bf x},{\bf C})+1).$$}\\

\begin{proof}
In Theorem 2.1, $h=\operatorname{rank}({\bf x}, {\bf C})$, $n-h-t=n-\operatorname{rank}({\bf x}, {\bf C})-t=L({\bf x})-1+n-S({\bf x})$. Then $t=S({\bf x})-L({\bf x})-\operatorname{rank}({\bf x}, {\bf C})+1$. The conclusion follows.
\end{proof}

{\bf Corollary 2.2.} {\em Let ${\bf C} \subset {\bf F}_q^n$ be a linear $[n,k]_q$ code with the minimum Hamming distance $d_H > \frac{n}{2}$. Suppose there exists a minimum Hamming weight codeword in ${\bf C}$ with consecutive index support. Then $$d_{insdel} \leq 2(d_H-k+1).$$}\\

\begin{proof}
First of all we have $k \leq d_H$, otherwise from $d_H <k \leq n-d_H+1$ we have $d_H \leq \frac{n}{2}$, which is contradict to the condition $d_H>\frac{n}{2}$. On the other hand there are $k$ linear independent columns among any $d_H$ columns. Otherwise we have a codeword with weight at most $n-d_H <d_H$ which is contradict to the condition $d_H >\frac{n}{2}$. The conclusion follows from Corollary 2.1 immediately.
\end{proof}

The new upper bound in Corollary 2.1 and 2.2 can be used to get some better upper bounds on the insdel distances of binary Reed-Muller codes and some algebraic geometric code with special coordinate orderings in the next section.\\

{\bf Corollary 2.3.} {\em Let ${\bf C} \subset {\bf F}_q^n$ be a linear $[n,k]_q$ code. If $d_H \geq k$, then its insdel distance satisfies $$d_{insdel} \leq 2(n-2k+2).$$ If $d_H\leq k-1$ then its insdel distance satisfies $d_{insdel} \leq 2(n-k-d_H+2)$. Hence we have $$d_{insdel} \leq \max\{2(n-2k+2),2(k-1)\}$$ for any $[n,k]_q$ linear code over ${\bf F}_q$. When $k \leq \frac{n+3}{3}$, we have $$d_{insdel} \leq 2(n-2k+2).$$
We also have $$d_{insdel} \leq \max\{2(n-2k+2),2(n-k-d_H+2)\}.$$}\\

\begin{proof}
If $d_H \geq k$ the last $n-k+1$ columns in any generator matrix of this code contain $k$ linear independent vectors in ${\bf F}_q^k$. Hence we can find an information free coordinate set of the cardinality $k$ located in $\{k,\ldots,n\}$. It is clear we can find a codeword such that the first $k-1$ coordinates are zero, then $d_{insdel} \leq 2(n-2k+2)$ follows from Theorem 2.1 for $h=k, t=n-2k+1$. Since $d_{insdel} \leq 2d_H$, we have $$d_{insdel} \leq \max\{2(n-2k+2),2k\}.$$

If $d_H \leq k-1$, in the generator matrix there are $k$ linear independent columns among the last $n-d_H+1$ columns, we can find an information free coordinate set of cardinality $k$ located in the coordinate position set $\{d_H,\ldots, n\}$. It is clear we can find a codeword with the first $d_H-1$ zero coordinates since $d_H-1 \leq k-2$. Then we have $$d_{insdel} \leq 2(n-k-d_H+2).$$ from Theorem 2.1 for $h=k$ and $t=n-k-d_H+1$.  From the direct Singleton upper bound $d_{insdel} \leq 2d_H$, then $d_{insdel} \leq \max \{2(n-2k+2), 2(k-1)\}$. When $k \leq \frac{n+3}{3}$, $2(k-1) \leq 2(n-2k+2)$, therefore $d_{insdel} \leq 2(n-2k+2)$. The conclusions follows immediately.
\end{proof}

Though the upper bound $d_{insdel} \leq 2(n-2k+2)$ follows from Theorem 2.1 in the case $k \leq \frac{n+3}{3}$. Actually this upper bound $d_{insdel} \leq 2(n-2k+2)$ is true for arbitrary linear codes. The following result and its proof is basically the same as \cite{AGFC,CGHL21}.  The half-Singleton bound in \cite{CGHL21} was proved from the result in \cite{AGFC}  by shortening. Our approach is more direct.\\

{\bf Half-Singleton bound (adapted from \cite{AGFC},\cite{CGHL21} Section5).} {\em Let ${\bf C}$ be a linear $[n,k]_q$ code satisfying $2k>n$ then there exists a non-zero codeword ${\bf x}=(x_1,\ldots,x_n)$ such that $(x_2,x_3,\ldots,x_n,x_1)$ is also a codeword in ${\bf C}$. Hence we have the half-Singleton bound $$d_{insdel} \leq \max\{2(n-2k+2),2\}.$$}

\begin{proof}
Let $H$ be the $(n-k) \times n$ parity-check matrix of this code ${\bf C}$ with $n$ columns ${\bf h}_1,\ldots,{\bf h}_n$. We form two new matrices as follows. One is the $(n-k) \times n$ matrix $H'=({\bf h}_2,{\bf h}_3,\ldots,{\bf h}_n,{\bf h}_1)$. Another is the $2(n-k) \times n$ matrix $H''$ by concatenation corresponding columns in $H$ and $H'$, that is, the $n$ columns in $H''$ are $n$ column vectors in ${\bf F}_q^{2(n-k)}$, $$({\bf h}_1,{\bf h}_2)^{\tau},({\bf h}_2,{\bf h}_3)^{\tau},\ldots,({\bf h}_{n-1},{\bf h}_n)^{\tau}, ({\bf h}_n, {\bf h}_1)^{\tau},$$ where ${\tau}$ is the transposition. Since $2(n-k)<n$, there is a non-zero solution of the equation $$H'' \cdot {\bf x}^{\tau}={\bf 0}.$$ This is the codeword claimed in the conclusion. By shorting $(n-2k+1)$ coordinates positions outside an information set, we get a linear $[2k-1,k]_q$ code with the insdel distance at most $2$. That is we have two codewords in this shortening code with a length $2k-2$ common subsequence. Then there are two codewords in the original code with the common  subsequence of the length at least $2k-2$.  The half-Singleton bound follows immediately.
\end{proof}

{\bf Corollary 2.4.} {\em We have  $$d_{insdel} \leq \inf_{1\leq r \leq k}\max\{2(d_r-2r+2),2\}$$ from the half-Singleton bound. Hence we have $$d_{insdel} \leq \inf_{1 \leq r \leq k} \max \{2([\frac{n(q^k-q^{k-r})}{q^k-1}]-2r+2),2\}.$$}\\

\begin{proof}
This is from the fact that the insdel distances of sub-codes of the code ${\bf C}$ is bigger than or equal to the insdel distance $d_{insdel}$ of this code ${\bf C}$. The first conclusion follows. The second upper bound follows from the Plotkin upper bound $d_r \leq [\frac{n(q^r-1)q^{k-r}}{q^k-1}]$ for the generalized Hamming weighs in \cite{TV1}.
\end{proof}

When $r=1$, $d_{insdel} \leq 2(d_H-2+2)=2d_H$, this is the direct upper bound on the insdel distances from the Hamming weight. When $r=k$, this is the half-Singleton bound. Thus Corollary 2.4 is a natural stronger generalization of these two previous known upper bounds.\\

For a linear MDS code, since the set of the first $k$ positions is an information free coordinate set, we always have $d_{insdel} \leq \max\{2(n-2k+2),2\}$ from Theorem 2.1. We conjecture that the upper bound $2(n-2k+2)$ can be attained for some Reed-Solomon codes in the first version of this paper \cite{Chen}. This conjecture was proved in a very recent paper \cite{CSI}.  The existence of Reed-Solomon codes with their insdel distances attaining the bound $2(n-2k+2)$ were proved for any dimension $k$. The code lengths of two dimension Reed-Solomon codes attaining the half-Singleton bound in \cite{CSI} are much longer than the code lengths in \cite{Duc21,CZ21}.\\

For a linear code the $r$-th generalized Hamming weight $d_r$ satisfies $d_r \leq n-k+r$, see \cite{Wei}. Thus if we combine this Singleton upper bounds for the generalized Hamming weights with the upper bounds in Corollary 2.4 directly,  the upper bound $2(d_r-2r+2)=2(n-k-r+2)$ is worse than the half-Singleton bound $2(n-2k+2)$ when $r<k$.\\

It is clear that for a linear $[n,k]_q$ code over ${\bf F}_q$ with the minimum Hamming weight $d_1$ and the 2nd generalized Hamming weight $d_2=d_1+1$ satisfying $d_1 < n-2k+3,$ then our bound $2(d_2-4+2)=2(d_1-1)$ is better than the direct bound $2d_1$ and the half-Singleton bound $2(n-2k+2)$.  Hence it is easy to construct linear codes over large fields to show that the half-Singleton bound and the direct bound  are not tight, though such linear codes are not natural.\\

\section{Discussion on coordinate-orderings}

We observe some examples of linear codes and show that the coordinate-orderings strongly affect the insdel distances of these linear codes.\\

Let ${\bf C}$ be an algebraic-geometric code over ${\bf F}_4$ defined by the Hermitian curve $y^2z+zy^2=x^3$ over ${\bf F}_4$, with the length $8$, the dimension $3$ and the minimum Hamming distance $5$. Let $\omega$ be the element in ${\bf F}_4$ such that $\omega^2+\omega+1=0$. Then the $8$ rational points of the above elliptic Hermitian curve is of the form $P_1=(0,0),P_2=(1,0),P_3=(\omega,1),P_4=(\omega,\omega),P_5=(\omega,\omega^2),P_6=(\omega^2,1),P_7=(\omega^2,\omega),P_8=(\omega^2,\omega^2)$. The above dimension $3$ algebraic-geometric code has one generator matrix of the following form.\\

$$
\left(
\begin{array}{cccccccc}
1&1&1&1&1&1&1&1\\
0&1&\omega&\omega&\omega&\omega^2&\omega^2&\omega^2\\
0&0&1&\omega&\omega^2&1&\omega&\omega^2\\
\end{array}
\right)
$$
From \cite{Munuera} the $2nd$ generalized Hamming weight is $d_2=7$. We observe that $d_1-2+2=d_1=5, d_2-4+2=5, d_3-6+2=4$, then the best coordinate ordering-free upper bound $8$ in this case is from the half-Singleton upper bound.\\

We now fix the ordering of coordinate positions as above. From Corollary 2.2 we have a better upper bound $d_{insdel} \leq 2(d_H-k+1)=6$ than the half-Singleton bound, since there is one weight $5$ codeword $(\omega^2,\omega,1,1,1,0,0,0)$ with consecutive index support. In this case the upper bound in Corollary 2.2 is better than the half-Singleton bound.\\

We consider the following two codewords ${\bf x}_1=(001 \omega \omega^2 1 \omega \omega^2)$ and ${\bf x}_2=(00 \omega \omega^2 1 \omega \omega^2 1)$. They have a common subsequence $(00\omega\omega^21\omega\omega^2)$ of length $7$. Hence the above two upper bounds are not tight for this Hermitian code. The insdel distance of this Hermitian code is $2$ with the above coordinate ordering.\\

We consider the following ordering of $8$ points $P_1,P_2,P_3,P_6,P_4,P_7,P_5,P_8$. The generator matrix is as follows.\\

$$
\left(
\begin{array}{cccccccc}
1&1&1&1&1&1&1&1\\
0&1&\omega&\omega^2&\omega&\omega^2&\omega&\omega^2\\
0&0&1&1&\omega&\omega&\omega^2&\omega^2\\
\end{array}
\right)
$$
The three columns at $(123)$, $(678)$ ,$(178)$, $(128)$ positions are linear independent. Then we do not have a weight $5$ codeword with consecutive index support. The best upper bound from Corollary 2.2 is $2(6-3+1)=8$, which is the same as the half-Singleton bound. There are two codewords $(0011\omega \omega \omega^2 \omega^2)$ and $(00\omega \omega \omega^2 \omega^2 11)$. Hence the insdel distance of this code with the above coordinate ordering is at most $4$.\\

The binary Reed-Muller codes are defined as follows. Let $P_1,\ldots,P_n$ be $n=2^m$ points of ${\bf F}_2^m$. Let $u\leq m$ be a positive integer. Set ${\bf Function}(u,m)$ be the set of linear combinations of monomials $x_{i_1} x_{i_2} \cdots x_{i_t}$, $t \leq u$. The dimension of ${\bf Function}(u,m)$ is $$1+m+\displaystyle{m \choose 2}+\cdots+\displaystyle{m \choose u}.$$ The binary Reed-Muller code $RM(u,m)$ is defined by $$RM(u,m)=\{(f(P_1),\ldots,f(P_n)):f \in {\bf Function}(u,m)\}.$$ The dimension is $$k=1+m+\displaystyle{m \choose 2}+\cdots+\displaystyle{m \choose u},$$  and the minimum distance is $$d_1=2^{m-u}.$$ The generalized Hamming weights of binary Reed-Muller codes were determined in \cite{Wei}. The insertion-deletion error-correcting capabilities of the first order binary Reed-Muller code was studied in \cite{DA}.  We can upper bound the insdel distances of binary Reed-Muller codes from our main result.  \\

We consider the 1st order binary Reed-Muller code, with the length $2^m$, the dimension $m+1$ and the minimum Hamming distance $2^{m-1}$. Since Reed-Muller codes are evaluation codes at $2^m$ points of ${\bf F}_2^m$, the coordinate positions are corresponding to $2^m$ points of ${\bf F}_2^m$. The supports of minimum weight codewords are affine subspaces of  ${\bf F}_2^m$.  Suppose that these coordinate positions are arranged as follows. The $2^{m-1}$ points in the linear subspace defined by $x_1=0$ and the affine subspace defined by $x_1=1$ are consecutive coordinate positions in its support. From Corollary 2.1 we have $d_{insdel} \leq 2(2^{m-1}-m)$, since there is an information free subset of $m$ points in the affine subspace defined by $x_1=1$.\\

From Corollary 2.1 we have the following upper bounds on insdel distances of binary Reed-Muller codes, which is dependent on the special ordering of coordinate positions.\\

{\bf Theorem  3.1.} {\em By arranging the coordinate positions corresponding to the points in the affine subspace defined by $x_1 \cdot x_2 \cdots x_u=1$ as consecutive index coordinate positions, the insdel distance of binary Reed-Muller code $RM(u,m)$ satisfying $u <\frac{m}{2}$ is at most  $2(1+\displaystyle{m-u \choose u+1}+\displaystyle{m-u \choose u+2}+\cdots+\displaystyle{m-u \choose m-u})$.}\\

\begin{proof}
Over the affine subspace defined by $x_1=x_2=\cdots=x_u=1$, the $1+\displaystyle{m-u \choose 1}+\cdots+\displaystyle{m-u \choose u}$ monomials $x_{j_1} \cdots x_{j_t}$, where $t \leq u$ and $j_1, \ldots, j_t \in \{u+1, \ldots,m\}$ are linear independent. This is the Reed-Muller code $RM(m-u, u)$. Then for a minimum weight codeword supported at this affine subspace, we have an information free subset with $1+\displaystyle{m-u \choose 1}+\cdots+\displaystyle{m-u \choose u}$ coordinate positions in its support. From Corollary 2.1 the conclusion follows.
\end{proof}

We consider the following case. Set $m=2m_1+1$ and $u=m_1-1$. Then the ordering-free upper bound from Corollary 2.2  is $$2(\displaystyle{2m_1+1 \choose m_1}+1).$$ However the ordering-depending upper bound from Theorem 3.1  is $$\frac{m_1^2+5m_1+8}{2}.$$ The Hamming distance of this Reed-Muller code $2^{2m_1+1-m_1+1}=2^{m_1+2}$ is exponential in $u$ and the insdel distance with respect to this special coordinate ordering is upper bounded by $poly(u)$ when $m$ goesto the infinity. Similarly set $u=m_1-c$, $c$ is a fixed positive integer, when $m_1$ goes to the infinity, from Theorem 3.1  the insdel distance of Reed-Muller code $RM(u,m)$ with respect to the special coordinate ordering is upper bounded by $poly(u)$ depending the positive integer $c$. The Hamming distances of these codes are exponentials of $u$. \\

\section{Insdel distances of  AGFC codes}

For a linear cyclic code, $(c_1,c_2\ldots,c_{n-1},c_n)$ and $(c_2,c_3,\ldots,c_n,c_1)$ are codewords, then their insdel distance is $2$. If the coordinate ordering is re-arranged, this is not true again. In \cite{AGFC} it was showed that by inserting one coordinate into codewords, cyclic codes can be used to correct at least one deletion. In this section we use our new bounds to give upper bounds of insdel distances of coordinate rearranged cyclic codes satisfying $k < \frac{n}{2}$. All these bounds are valid for any rearranged coordinate ordering of this cyclic codes.\\

In \cite{YF} many cyclic $[n,k]$ codes ${\bf C}_{n,k}$ over ${\bf F}_q$ with length $n=\frac{q^k-1}{e}$ and $d_r=\frac{n(q^k-q^{k-r})}{q^k-1}$, $1\leq r \leq k$, were constructed. The generalized Hamming weights of these codes attain the Plotkin bound $d_r=[\frac{n(q^r-1)q^{k-r}}{q^k-1}]$, see \cite{TV1}. Denote the code with the  rearranging the coordinate ordering of ${\bf C}_{n,k}$ by ${\bf C}_{n,k,rearranged}$. In \cite{AGFC} new linear codes were constructed by the following inserting coordinate construction from a binary cyclic code ${\bf C} \subset {\bf F}_2^n$ . Let $f: {\bf F}_2^n \longrightarrow {\bf F}_2$ be defined by $f((c_1,\ldots,c_n))=c_1$, if $c_1=c_2=\cdots=c_n$ or $f((c_1,\ldots,c_n))=c_{\lfloor\frac{n}{2}\rfloor+1}$ otherwise. The new code ${\bf C}_{f,\lfloor\frac{n}{2}\rfloor}$ is a length $n+1$ code over ${\bf F}_2$ by inserting $f({\bf c})$ at the $\lfloor\frac{n}{2}\rfloor$ position of all codewords in ${\bf C}$. This is a linear binary codes since $f$ is a linear mapping. It was proved in \cite{AGFC} that the linear code ${\bf C}_{f,\lfloor\frac{n}{2}\rfloor}$ can correct at least one deletion. \\

We consider the following construction. Let $f$ be any non-trivial linear function on ${\bf F}_q^n$ and ${\bf C}_{f,h}$ be the new linear code of length $n+1$ consisting of all codewords by inserting $f({\bf c})$ at the $h$-th coordinate position of all codewords ${\bf c}$ in ${\bf C}$. We call this linear code AGFC code. The coordinate ordering rearranged linear code ${\bf C}_{f,h}$ of such code from ${\bf C}_{n,k}$ in \cite{YF} and $f$ is denoted by ${\bf C}_{f,h,rearranged}$. Then it is clear $d_r({\bf C}_{f,h,rearranged}) \leq d_r({\bf C})+1$.\\

From our new upper bounds based on the generalized Hamming weights, we can get upper bounds $$d_{insdel}({\bf C}_{f,h,rearranged}) \leq \inf_{1\leq i \leq k}\max\{2(d_r({\bf C})-2r+3),2\}.$$  We consider coordinate rearranged AGFC codes ${\bf C}_{n,k,f,h,raarranged}$ from these cyclic codes ${\bf C}_{n,k}$ in \cite{YF}. Then the following coordinate-ordering-free upper bounds for cyclic codes and related AGFC codes with any coordinate ordering follow from Corollary 2.4.\\

{\bf Proposition 4.1.} {\em We have $d_{insdel}({\bf C}_{n,k,rearranged}) \leq 2(\frac{n(q^k-q^{k-r})}{q^k-1}-2r+2),$  and $d_{insdel}({\bf C}_{n,k, f,h,rearranged}) \leq 2(\frac{n(q^k-q^{k-r})}{q^k-1}-2r+3)$ for $1 \leq r \leq k$.}\\

\section{Conclusion and open problems}

We give new coordinate-ordering-free upper bounds on the insdel distances of linear codes, which are stronger than some previous known bounds.  They are applied to one algebraic-geometric code from the Hermitian curve over ${\bf F}_4$, some Reed-Muller codes and some AGFC codes. It seems that insdel distances of linear codes are easy to be upper bounded, but very hard to be lower bounded. The insdel distances of linear codes keep mysterious as in the following problems.\\

1) Are upper bounds in Theorem 2.1 and Corollary 2.4 tight for general linear $[n,k]_q$ code?\\

2) If the answer to the problem 1) is positive, can these optimal linear codes attaining these bounds be explicitly constructed? What is the longest possible lengths of these optimal linear codes attaining these upper bounds? We refer to \cite{CSI} for the latest existence results about Reed-Solomon codes attaining the half-Singleton bound.\\

More importantly we need some good lower bounds on insdel distances of linear codes over small fields.\\

3) Can some good lower bounds on the insdel distances be established for some well-constructed binary linear codes? Or is there a nice coordinate ordering such that the insdel distance $d_{inedel}$ of a given binary linear code can be lowered bounded directly from the Hamming distances $d_H$?\\

In our recent paper \cite{Chen1} subspace-metric and subset-metric codes were introduced and constructed. The minimum subspace distances and the minimum subset distances of codes are natural lower bounds for the minimum insdel distances. However most subspace-metric codes and subset-metric codes in \cite{Chen1} are defined over large fields and not linear. It seems that lower bounding insdel distances of linear codes over small fields is a difficulty problem.\\

{\bf Acknowledgement.} The author thanks Dr. Shu Liu and Professor Bocong Chen for introducing me to the topic of insertion-deletion codes. The author is grateful to Professor B. Haeupler for his very helpful comment and criticism on the 1st version of this paper. The author thanks two referees and the Associate Editor sincerely for their suggestions to improve the presentation of the paper.\\

\ifCLASSOPTIONcaptionsoff
  \newpage
\fi



%

%

\begin{IEEEbiography}{Hao Chen}
Hao Chen obtained his PH.D degree in mathematics in the Institute of Mathematics, Fudan University in 1991. He is now a professor of the College of Information Science and Technology/Cyber Security, Jinan University. His research interests are coding and cryptography, quantum information and computation, lattices and  algebraic geometry.
\end{IEEEbiography}




\end{document}